\documentclass[12pt]{article}

\usepackage{amsmath,amsfonts,amssymb}

\usepackage{subfigure}

\usepackage{epsfig,graphicx,bm}

\usepackage{epstopdf}

\usepackage{lipsum}

\begin{document}

\begin{titlepage}

\begin{center}

\vskip 0.4 cm

\begin{center}

{\Large \bf Cosmological Perturbations in Restricted

$f(R)$-Gravity}

\end{center}

\vskip 1cm

\vspace{1em} M. Chaichian,$^a$ A. Ghalee,$^b$ J. Kluso\v{n},$^{c,}$



\footnote{Email addresses:

masud.chaichian@helsinki.fi   (M. Chaichian),ghalee@ut.ac.ir   (A.
Ghalee), klu@physics.muni.cz   (J.Kluso\v{n}), }\\

\vspace{1em}$^a$\textit{Department of Physics, University of
Helsinki,P.O. Box 64,\\ FI-00014 Helsinki, Finland}\\
\vspace{.3em} $^b$\textit{Department of Physics, Tafresh University,
Tafresh 39518 79611, Iran}\\
\vspace{.3em} $^c$\textit{Department of Theoretical Physics and
Astrophysics, Faculty of Science,\\
Masaryk University, Kotl\'a\v{r}sk\'a 2, 611 37, Brno, Czech
Republic}

\vskip 0.8cm

\end{center}

\begin{abstract}

We investigate the metric perturbations of the restricted $f(R)$
theory of gravity in the cosmological context and  explore the
phenomenological implications of this model. We show that it is
possible to construct a restricted model of gravity, in which the
background equations are the same as the equations of motion which
are derived from the Einstein-Hilbert action with the cosmological
constant term. We argue that the deviation from the Einstein-Hilbert
model emerges in the perturbed equations, for which we have a
non-vanishing anisotropic stress. Further, with the help of the
results of Planck data for the modified gravity we obtain
constraints on the parameters of the model.
\end{abstract}
\end{titlepage}

\bigskip

\newpage

\def\bn{\mathbf{n}}

\newcommand{\bC}{\mathbf{C}}

\newcommand{\bD}{\mathbf{D}}

\def\hf{\hat{f}}

\def\tK{\tilde{K}}

\def\tmG{\tilde{\mG}}

\def\mC{\mathcal{C}}

\def\bk{\mathbf{k}}

\def\tpi{\tilde{\pi}}

\def\tp{\tilde{p}}

\def\tr{\mathrm{tr}\, }

\def\tmH{\tilde{\mH}}

\def\tf{\tilde{f}}

\def\tY{\mathcal{Y}}

\def\nn{\nonumber \\}

\def\bI{\mathbf{I}}

\def\tmV{\tilde{\mV}}

\def\e{\mathrm{e}}

\def\bE{\mathbf{E}}

\def\bX{\mathbf{X}}

\def\bY{\mathbf{Y}}

\def\bR{\bar{R}}

\def\hN{\hat{N}}

\def\hK{\hat{K}}

\def\hnabla{\hat{\nabla}}

\def\hc{\hat{c}}

\def\mH{\mathcal{H}}

\def\hZ{\hat{Z}}

\def\bz{\mathbf{z}}

\def\bK{\mathbf{K}}

\def\tmJ{\tilde{\mathcal{J}}}

\def\tr{\mathrm{Tr}}

\def\mJ{\mathcal{J}}

\def\partt{\partial_t}

\def\parts{\partial_\sigma}

\def\bG{\mathbf{G}}

\def\str{\mathrm{Str}}

\def\Pf{\mathrm{Pf}}

\def\bM{\mathbf{M}}

\def\tA{\tilde{A}}

\newcommand{\mW}{\mathcal{W}}

\def\bx{\mathbf{x}}

\def\by{\mathbf{y}}

\def \mD{\mathcal{D}}

\newcommand{\tZ}{\tilde{Z}}

\newcommand{\tW}{\tilde{W}}

\newcommand{\tmD}{\tilde{\mathcal{D}}}

\newcommand{\tN}{\tilde{N}}

\newcommand{\hC}{\hat{C}}

\newcommand{\hg}{g}

\newcommand{\hX}{\hat{X}}

\newcommand{\bQ}{\mathbf{Q}}

\newcommand{\hd}{\hat{d}}

\newcommand{\tX}{\tilde{X}}

\newcommand{\calg}{\mathcal{G}}

\newcommand{\calgi}{\left(\calg^{-1}\right)}

\newcommand{\hsigma}{\hat{\sigma}}

\newcommand{\hx}{\hat{x}}

\newcommand{\tchi}{\tilde{\chi}}

\newcommand{\mA}{\mathcal{A}}

\newcommand{\ha}{\hat{a}}

\newcommand{\tB}{\tilde{B}}

\newcommand{\hrho}{\hat{\rho}}

\newcommand{\hh}{\hat{h}}

\newcommand{\homega}{\hat{\omega}}

\newcommand{\mK}{\mathcal{K}}

\newcommand{\hmK}{\hat{\mK}}

\newcommand{\hA}{\hat{A}}

\newcommand{\mF}{\mathcal{F}}

\newcommand{\hmF}{\hat{\mF}}

\newcommand{\hQ}{\hat{Q}}

\newcommand{\mU}{\mathcal{U}}

\newcommand{\hPhi}{\hat{\Phi}}

\newcommand{\hPi}{\hat{\Pi}}

\newcommand{\hD}{\hat{D}}

\newcommand{\hb}{\hat{b}}

\def\I{\mathbf{I}}

\def\tW{\tilde{W}}

\newcommand{\tD}{\tilde{D}}

\newcommand{\mG}{\mathcal{G}}

\def\IT{\I_{\Phi,\Phi',T}}

\def \cit{\IT^{\dag}}

\newcommand{\hk}{\hat{k}}

\def \cdt{\overline{\tilde{D}T}}

\def \dt{\tilde{D}T}

\def\bra #1{\left<#1\right|}

\def\ket #1{\left|#1\right>}

\def\mV{\mathcal{V}}

\def\Xn #1{X^{(#1)}}

\newcommand{\Xni}[2] {X^{(#1)#2}}

\newcommand{\bT}{\mathbf{T}}

\def\bmR{\bar{\mR}}

\newcommand{\mL}{\mathcal{L}}

\newcommand{\mbQ}{\mathbf{Q}}

\def\mat{\tilde{\mathbf{a}}}

\def\mtF{\tilde{\mathcal{F}}}

\def \tZ{\tilde{Z}}

\def\mtC{\tilde{C}}

\def \tY{\tilde{Y}}

\def\pb #1{\left\{#1\right\}}

\newcommand{\E}[3]{E_{(#1)#2}^{ \quad #3}}

\newcommand{\p}[1]{p_{(#1)}}

\newcommand{\hEn}[3]{\hat{E}_{(#1)#2}^{ \quad #3}}

\def\mbPhi{\mathbf{\Phi}}

\def\tg{\tilde{g}}

\newcommand{\phys}{\mathrm{phys}}

\def\tmC{\tilde{\mC}}

\section{Introduction}
Recently a new model of restricted $f(R)$-gravity has been proposed
in \cite{Chaichian:2015qjo}. This proposal is based on  the idea of
a mild breaking of the diffeomorphism invariance of the four-
dimensional $f(R)$ gravity. Recall that $f(R/M^2_P)$-gravity is
described by the action \footnote{For a review and extensive list of
references, see \cite{DeFelice:2010aj}.}
\begin{equation}\label{i-0}
S_{f(R)}=\int d^{4}x\sqrt{-g} \left[\frac{M_{P}^{2}R}{2}+\eta
M_{P}^{4}f\left(\frac{R}{M_{P}^{2}}\right)\right] +S_{Matter} \ ,
\end{equation}
where $M_{P}^{2}$ is the reduced Planck mass and $\eta$ is a
dimensionless constant. This action is invariant under the full
four-dimensional diffeomorphism by construction. In order to find
the restricted form of $f(R)$-gravity, it is useful to introduce the
$3+1$-decomposition of the metric $g_{\mu\nu}$
\cite{Arnowitt:1962hi,Gourgoulhon:2007ue}
\begin{eqnarray}\label{hgdef}
g_{00}=-N^2+N_i h^{ij}N_j , \quad g_{0i}=N_i , \quad
g_{ij}=h_{ij} \ ,\nonumber \\
g^{00}=-\frac{1}{N^2} , \quad g^{0i}=\frac{N^i}{N^2} , \quad
g^{ij}=h^{ij}-\frac{N^i N^j}{N^2} \ ,
\end{eqnarray}
where we have defined $h^{ij}$ as the inverse of the induced metric
$h_{ij}$ on the Cauchy surface $\Sigma_t$ at each time $t$,
\begin{equation}
h_{ik}h^{kj}=\delta_i^{ \ j}
\end{equation}
and we denote $N^i=h^{ij}N_j$. The four-dimensional scalar curvature
in $3+1$ formalism has the form
\begin{eqnarray}
R(g_{\mu\nu})
&=&K_{ij}\mG^{ijkl}K_{kl}+{}^{(3)}R+\frac{2}{\sqrt{-g}}\partial_\mu(\sqrt{-g}
n^\mu K) -\frac{2}{\sqrt{h}N}\partial_i(\sqrt{h}h^{ij}\partial_j N)
\nonumber \\
&\equiv &K_{ij}\mG^{ijkl}K_{kl}+{}^{(3)}R+\Xi \ , \nonumber \\
\end{eqnarray}
where the extrinsic curvature of the spatial hypersurface $\Sigma_t$
at the time $t$ is defined as
\begin{equation}
K_{ij}=\frac{1}{2N}\left(\frac{\partial h_{ij}}{\partial t}- D_i
N_j-D_j N_i\right) ,
\end{equation}
with $D_i$ being the covariant derivative determined by the metric
$h_{ij}$, and where the de Witt metric is defined as
\begin{equation}
\mG^{ijkl}=\frac{1}{2}(h^{ik}h^{jl}+h^{il}h^{jk}) -h^{ij}h^{kl}
\end{equation}
with inverse
\begin{equation}
\mG_{ijkl}=\frac{1}{2}(h_{ik}h_{jl}+h_{il}h_{jk}) -\frac{1}{2}
h_{ij}h_{kl} \
\end{equation}
which obeys the relation
\begin{equation}
\mG_{ijkl}\mG^{klmn}=\frac{1}{2}(\delta_i^m\delta_j^n+\delta_i^n
\delta_j^m) \ .
\end{equation}
Further, $n^\mu$ is the future-pointing unit normal vector to the
hypersurface $\Sigma_t$, which is written in terms of the ADM
variables as
\begin{equation}
n^0=\sqrt{-g^{00}}=\frac{1}{N} ,\quad
n^i=-\frac{g^{0i}}{\sqrt{-g^{00}}}=-\frac{N^i}{N} \ .
\end{equation}
In order to formulate the restricted $f(R)$-gravity we break the
full diffeomorphism invariance of the $f(R)$-gravity by  performing
the replacement
\begin{equation}\label{i-2}
R\rightarrow R+(\Upsilon-1)\Xi \ ,
\end{equation}
where $\Upsilon$ is a dimensionless parameter that controls the
breaking of the diffeomorphism invariance of the action. Even if the
shift (\ref{i-2}) seems to be very mild,  it turns out that it has
significant impact on the Hamiltonian structure of this theory as
was shown in \cite{Chaichian:2015qjo}. Careful analysis performed
there showed that in order to have a consistent theory from the
Hamiltonian analysis point of view,  it is necessary to include
terms which depend on the spatial derivative of the lapse
$a_i=\frac{\partial_i N}{N}$ into the action. Further, the breaking
of the diffeomorphism invariance suggests the possibility to define
theory with the generalized de Witt metric
\cite{Horava:2008ih,Horava:2009uw}
\begin{eqnarray}
\tmG^{ijkl}&=&\frac{1}{2}(h^{ik}h^{jl}+h^{il}h^{jk})-\lambda
h^{ij}h^{kl} \ , \quad \lambda \neq \frac{1}{3} \ , \nonumber \\
\tmG_{ijkl}&=&\frac{1}{2}(h_{ik}h_{jl}+h_{il}h_{jk})-\frac{\lambda}{3\lambda-1}
h_{ij}h_{kl} \ . \nonumber \\
\end{eqnarray}
In summary, we proposed in \cite{Chaichian:2015qjo} the extended
form of the restricted $f(R)$-gravity when we performed the
replacement
\begin{equation}\label{repla}
R\rightarrow R_{\Upsilon}\equiv
K_{ij}\tmG^{ijkl}K_{kl}+{}^{(3)}R+\Upsilon\Xi +\gamma_1 a_i
a^i+\gamma_2 {}^{(3)}R^{ij}a_i a_j \ ,
\end{equation}
where $\gamma_1,\gamma_2$ are the  corresponding coupling constants.
We showed there that this theory is consistent from the Hamiltonian
analysis point of view since the structure of constraints is the
same as in case of non-projectable Ho\v{r}ava-Lifshitz gravity
\cite{Blas:2009qj,Blas:2010hb,Blas:2009ck}, for the Hamiltonian
analysis, see
\cite{Kluson:2010nf,Donnelly:2011df,Mukohyama:2015gia,Chaichian:2015asa}
and also
\cite{Kluson:2009rk,Kluson:2010za,Kluson:2010xx,Kluson:2009xx,
Carloni:2010nx,Chaichian:2010yi}. We also analyzed cosmological
solutions of the restricted $f(R)$-gravity and we found new
solutions whose properties depend on the value of the parameter
$\Upsilon$. It is important to emphasize  that some of these
solutions cannot be found in the diffeomorphism invariant
$f(R)$-gravity.

Due to the fact that there are new cosmological solutions, it is
natural to investigate them in more details. The aim of this paper
is to focus on the analysis of the cosmological fluctuations of the
restricted $f(R)$-gravity. As the first step we determine the
background equations from the restricted $f(R)$-gravity action when
we focus on i a time-dependent ansatz. Then we analyze the
fluctuations above this background. Due to the fact that the scalar,
vector and tensor fluctuations decouple at the quadratic order
\footnote{For review of cosmological fluctuations, see for example
\cite{mukhanov}.} we can analyze each kind of the fluctuations
separately. We show that the restricted $f(R)$-gravity differs from
the standard $f(R)$-gravity in the scalar sector, while the vector
and tensor sectors have the same properties as in the case of
diffeomorphism invariant $f(R)$-gravity. More precisely, we show
that with suitable choice of parameters the restricted
$f(R)$-gravity allows to explain recent observation data
 that predict the possibility of the existence of a non-zero
anisotropic stress (See for example \cite{silk}.).

This paper is organized as follows. In the next section
(\ref{second}) we briefly review the equations of motion of the
restricted $f(R)$-gravity evaluated on the  time-dependent
cosmological solution. In section (\ref{third}) we analyze the
fluctuations above this background and in section (\ref{fourth}) we
discuss our results with relation to the recent phenomenological
observations. Finally, in conclusion (\ref{fifth}) we outline our
results.
\section{ Background Equations}\label{second}
In this section we derive the equations of motion for the restricted
$f(R)$-gravity when we presume the Friedmann-Robertson-Walker (FRW)
form of the background. Recall that the restricted
$f(R/M^{2}_P)$-gravity action has the form
\begin{equation}\label{ac1}
S_{res}=\int d^{4}x\sqrt{-g} \left[M_{P}^{2}\frac{R_{\Upsilon}}{2}+
\eta
M_{P}^{4}f\big(\frac{R_{\Upsilon}}{M_{P}^{2}})\right]+S_{Matter} \ ,
\end{equation}
where $R_\Upsilon$ is defined in (\ref{repla}). The matter part
contributes to the gravitational equations of motion through the
stress-energy tensor which  is defined as
\begin{equation}
T_{\mu\nu}=-\frac{2}{\sqrt{-g}}\frac{\delta S_{Matter}}{\delta
g^{\mu\nu}} \ .
\end{equation}
Our goal is to analyze the spatially homogeneous and isotropic
Universe so that the metric ansatz is the FRW metric which  has the
form
\begin{equation}\label{1-1}
ds^{2}=-N(t)^{2}dt^{2}+a(t)^{2}dx^{i}dx^{j}\delta_{ij} \ ,
\end{equation}
where $N=N(t)$ is the lapse and $a=a(t)$ is the scale factor. As
usual, the Hubble parameter is defined as $H\equiv\dot{a}/a$. Note
that we cannot set $N(t)=1$ from the beginning due to the restricted
form of the diffeomorphism. But, since for the background equations
all quantities depend only on time, one can use redefinition of time
in \eqref{1-1} to set $N=1$ in the background equations. We also
presume matter in the form of a  prefect fluid which means that the
stress-energy tensor has the form
\begin{equation}
T^\alpha_{ \ \beta}=\left(\begin{array}{cccc}
-\rho(t) & 0 & 0 & 0 \\
0 & p(t) & 0 & 0 \\
0 & 0 & p(t) & 0 \\
0 & 0 & 0 & p(t) \\ \end{array}\right) \ .
\end{equation}
Since we presume that the full diffeomorphism is broken in the
gravitational sector only we find that the matter action is
diffeomorphism invariant. As a result the stress-energy tensor is
conserved in the sense  $\nabla_\mu T^{\mu\nu}=0$   on  the
condition that the matter satisfies the equation of motion. Now for
the background (\ref{1-1}) we obtain
\begin{eqnarray}
\Gamma^0_{00}=\frac{\dot{N}}{N} \ , \quad
\Gamma^0_{ij}=\frac{1}{N^2}\dot{a}a\delta_{ij} \ , \quad
\Gamma^0_{0i}=0 \ , \quad \Gamma^i_{00}=0 \ , \quad
\Gamma^i_{j0}=\delta^i_j H \ , \quad
\Gamma^i_{jk}=0 \ \nonumber \\
\nonumber \\
\end{eqnarray}
so that the conservation of the stress-energy tensor implies the
standard conservation equation
\begin{eqnarray}\label{consten}
\dot{\rho}+3H(\rho+p)=0 \ . \nonumber \\
\end{eqnarray}
Using (\ref{1-1}) and the general relations in the ADM formalism,
which are discussed in Ref. \cite{Gourgoulhon:2007ue}, one can show
that
\begin{equation}\label{i-3}
\Xi=-6\frac{H\dot{N}}{N^{3}}+6\frac{\dot{H}}{N^{2}}+18
\frac{H^{2}}{N^{2}}\hspace{.08cm}.
\end{equation}
The generalized Friedmman equation for the model has been determined
in \cite{Chaichian:2015qjo} as
\begin{equation}\label{gen11}
3AH^{2}+\eta M_{P}^{2}f+\eta (6AH^{2}-R_{\Upsilon}) F+6\eta\Upsilon
H \dot{F}=\frac{\rho}{M_{P}^{2}} \ ,
\end{equation}
where
\begin{equation}\label{fdef}
R_{\Upsilon}= -6AH^{2}+\Upsilon (6\dot{H}+18H^{2}), \hspace{0.15cm}
f\equiv f\big(\frac{R_{\Upsilon}}{M_{P}^{2}}\big)
\hspace{0.15cm},F\equiv M_{P}^{2}\frac{df}{dR_{\Upsilon}} \equiv
f'\hspace{0.15cm},A\equiv\frac{3\lambda-1}{2} \ .
\end{equation}
Note that we have defined prime as derivative with respect to the
argument of $f$. As a result, $F,f'',...$ are dimensionless
quantities. Finally, we  note that there is still another equation
after performing the time derivative of eq. \eqref{gen11} and then
use eq. \eqref{i-3}
\begin{equation}\label{bae}
A\dot{H}(1+2\eta F)+ \eta H (2A-3\Upsilon)\dot{F}+ \Upsilon
\eta\ddot{F}=-\frac{(\rho+p)}{2M_{P}^{2}} \ .
\end{equation}
After this brief review of the background equations we now switch to
the main topic of this paper which is the analysis of the
cosmological perturbations.
\section{ Cosmological Perturbations}\label{third}
The goal of this section is to derive equations for the perturbed
FRW space-time in the Arnowitt-Deser-Misner(ADM) formalism
\footnote{For a review, see \cite{Gourgoulhon:2007ue}}. As a check
of the validity of our approach, we note that for
$\Upsilon=1,\lambda=1, \gamma_{1}$ and $\gamma_{2}=0$ the results
derived in this paper should reduce to the corresponding results of
the standard $f(R/M_{P}^2)$ gravity, see for example the review
\cite{DeFelice:2010aj}.

To begin with,  we emphasize that we are writing our equations in
the Newtonian gauge which is defined by $N^{i}=0$. We use the
Newtonian gauge for the following reason: As it has been argued in
\cite{mukhanov}, one of the advantages of the Newtonian gauge is
that the physical fields which are defined by the gauge fixing,
coincide with the gauge-invariant variables. Note also that it is an
easy task to move from the Newtonian gauge to other gauges
\cite{mukhanov}.
Finally, note that due to the fact that the spatial section of the
metric (\ref{1-1}) is flat, it is natural to use the Fourier
decomposition of the perturbation where the Fourier components of a
general perturbation $U(t,\textbf{x})$ are defined as
\begin{equation}
U_k(t)=\int d^{3}\bx U(t,\textbf{x})e^{-i\textbf{k.x}} \ ,
\end{equation}
where $\bx=(x^1,x^2,x^3),\bk=(k^1,k^2,k^3)$. Further, we also
decompose $F$ and $\dot{F}$ into the homogeneous and perturbed parts
as
\begin{equation}
F=\bar{F}+\delta F,\hspace{.15cm} \dot{F}=\dot{\bar{F}}+\dot{\delta
F} \ ,
\end{equation}
where $\hspace{0.1cm}\bar{ }\hspace{0.1cm}$ over any quantity shows
the  unperturbed part of that quantity. In the case of the
fluctuations of the matter we use the following parameterization for
the perturbed energy-momentum tensor
\begin{equation}\label{fl1}
\delta T_{0}^{0}=-\delta \rho, \hspace{.1cm}\delta
T_{0}^{i}=-(\rho+p)\partial_{i}v, \hspace{.1cm}\delta
T_{i}^{j}=\delta p\delta_{i}^{j} \ ,
\end{equation}
where $v$ is the potential for the spatial velocity of the fluid.\\
\subsection{ The Scalar Metric Perturbations}
Before we proceed to the study of the perturbed equations it is
instructive to discuss the
role of diffeomorphism symmetry in the cosmological context.\\
For a full diffeomorphism invariant model, as for example the
Einstein-Hilbert action or $f(R)$gravity action, there is no a
priori preferred coordinate system. Of course the symmetry can be
broken by imposing other restrictions on the model. For example, in
the cosmological context, it has been shown that the only coordinate
system(observer) that is compatible with the assumption of
homogeneity and isotropy of space is described by the FRW metric
\cite{Weinbergbook}. The uniqueness of FRW metric helps us to choose
it as the background metric for our model in this paper.\ But, the
uniqueness of the metric is broken  if we consider  Universe which
is not homogeneous. The deviation from homogeneity is considered as
the perturbed FRW metric.
For example, the scalar metric perturbations can be parameterized as
\cite{mukhanov}
\begin{align}\label{note-add}
ds^{2}=-(1+2Y)dt^{2}+2a\partial_{i}B dx^{i}dt
+a^{2}[(1+2\zeta)\delta_{ij}+2\partial_{i}\partial_{j}
E]dx^{i}dx^{j} \ ,
\end{align}
where $Y=Y(t,\bx)$,$B=B(t,\bx)$,
$E=E(t,\bx)$, $\zeta=\zeta(t,\bx)$ are $3-$scalars.\\
It is an unnecessary and senseless task to insert the above
expression into general equations of motion in order to obtain the
equations of motion for these $3-$scalars. The reason for this
statement is that, similar to the model with the full diffeomorphism
symmetry, here we confront with the so-called gauge problem. The
problem arises when we note that the model is invariant under the
coordinate transformation as $t\rightarrow t+ P(t)$, $
x^{i}\rightarrow x^{i}+\partial^{i}K(t,\bx)$, where
$\partial^i=\delta^{ij}\partial_j$. Under this transformation the
components of the metric (\ref{note-add}) transform as
\cite{mukhanov}
\begin{eqnarray}\label{note-add1}
Y\rightarrow Y^{t}&=& Y-\dot{P}(t) \ ,\nonumber\\
B\rightarrow B^{t}&=& B+\frac{P(t)}{a}-a\dot{K}(t,\bx) \ ,\nonumber\\
E\rightarrow E^{t}&=&E-K(t,\bx) \ ,\nonumber\\
\zeta\rightarrow \zeta^{t}&=&\zeta-HP(t) \ .\nonumber\\
\end{eqnarray}
Therefore, even if we tried to obtain the equations for the
components of eq.\eqref{note-add}, we would have some solutions
which are unphysical in the sense that they can be derived by
application of the transformation (\ref{note-add1}) on some
particular solutions. In order to avoid the above problem, i.e. the
gauge problem, one can choose a specific gauge (coordinate) or use
the gauge-invariant quantities. In the process of the gauge fixing,
values for $P(t) $ and $ K(t,\bx) $ are explicitly specified such
that we have not any residual symmetry for the solution.
\textit{Physically, gauge fixing means that a
specific spatial coordinate has been chosen}.\\
In the following we will work with the Newtonian gauge. The
Newtonian gauge is defined as the perturbed metric for observers who
are immobile at the hypersurfaces of constant time. Also, the normal
vector of the hypersurfaces are the same as the worldlines of the
observers at any time. Thus, the Newtonian gauge is described by
\cite{mukhanov}
\begin{align}\label{note-add2}
ds^{2}=-(1+2Y)dt^{2}+a^{2}[(1+2\zeta)\delta_{ij}]dx^{i}dx^{j} \ .
\end{align}
The Newtonian gauge is the preferred gauge for the late-time
cosmology as was argued in \cite{Ade2:2015lrj}. Therefore, we will
use it to compare the results of our model with the Plank
observations. Finally we should
  emphasize that it is still
possible to construct the gauge invariant
quantities form the Newtonian gauge variables \cite{mukhanov}.\\
 To proceed
further, we note that it is convenient to parameterize the scalar
metric perturbations in the Newtonian gauge as
\begin{equation}\label{defNfluc}
N^{2}=1+2Y\equiv e^{2\Phi}, \hspace{.15cm} h_{ij}=a^{2}(1+2\zeta)
\delta_{ij}\equiv a^{2}e^{-2\Psi}\delta_{ij} \ ,
\end{equation}
where $\Phi$ and $\Psi$ are space and time dependent. Using these
definitions in (\ref{repla}) we easily find
\begin{equation}\label{Rper}
\begin{split}
R_{\Upsilon}|_{scalar}=&{}^{3}R+
6(3\Upsilon-A)e^{-2\Phi}(H-\dot{\Psi})^{2} +6\Upsilon
e^{-2\Phi}(\dot{H}-
\ddot{\Psi})\\
&-6\Upsilon\dot{\Phi}e^{-2\Phi} (H-\dot{\Psi})
-2\Upsilon\frac{e^{2\Psi}}{a^{2}}\partial^{2}\Phi\\
&+2\Upsilon\frac{e^{2\Psi}}{a^{2}}
\partial_{i}\Phi\partial_{i}\Psi+
(\gamma_{1}-2\Upsilon) \frac{e^{2\Psi}}{a^{2}}
\partial_{i}\Phi\partial_{i}\Phi \ .
\end{split}
\end{equation}
As a check, we note that for $\Phi=\Psi=0$ the above relation
reduces to the background form of $ R_{\Upsilon} $ given in
eq.\eqref{fdef}. Now by performing the linearization of the
expression (\ref{Rper}), we obtain Fourier component of $\delta
R_{\Upsilon}|_{scalar}$
\begin{equation}\label{ldr}
\begin{split}
\delta R_{\Upsilon}|_{scalar}(k)=&\frac{-4k^2}{a^{2}}\Psi
+\frac{2\Upsilon k^{2}}{a^{2}}\Phi-12(3\Upsilon-A)H^{2}\Phi\\
&-12\Upsilon \dot{H}\Phi-6\Upsilon
H\dot{\Phi}-6\Upsilon\ddot{\Psi}-12(3\Upsilon-1)H \dot{\Psi} \ ,
\end{split}
\end{equation}
where $k^2\equiv k_i k^i$. Further, inserting (\ref{fl1}) and
(\ref{defNfluc}) into the equation (\ref{consten}) and by performing
the corresponding linearization, we obtain two equations
\begin{equation}\label{im1}
\dot{\delta\rho}+3H(\delta\rho+\delta p)=\frac{k^2}{a^2}\delta q
+3(\rho+p)\dot{\Psi} \ ,
\end{equation}
and
\begin{equation}\label{im2}
\dot{\delta q}+3H\delta q+\delta p+(\rho+p){\Phi}=0 \ ,
\end{equation}
where $\delta q\equiv-(\rho+p)v$. Note that these equations have the
same forms as the corresponding equations for the usual $f(R/M^{2})$
gravity. In order to derive the remaining equations, we proceed in
the following way. To begin with, we insert \eqref{Rper} into the
action. As we pointed out we work in the Newtonian gauge. Then in
order to vary the action with respect to the shift, it is sufficient
to consider the terms which are proportional to $N_{i}\Psi$ and
$N_{i}\Phi$. For example, if we define $\delta_{N_{i}}$ as a
variation with respect to the shift then for $S_{Matter}$ we obtain
\begin{equation}
\delta_{N_{i}} S_{Matter}=- \frac{1}{2}\int
d^{4}x\sqrt{-g}T^{\mu\nu}\delta_{N_{i}} g_{\mu\nu}=-\frac{1}{2}\int
d^{4}x a^{3}T^{0i}\delta N_{i}+\mathcal{O}(N_i^{2}) \ .
\end{equation}
Using this result and also using the following relations
\begin{equation}
n^{\mu}=\big(e^{-\Phi},-N^{i}e^{-\Phi}\big),\hspace{.1cm}
\Gamma^i_{ij}=-3\partial_{j}\Psi
\end{equation}
and after some integration by parts we obtain
\begin{equation}
\begin{split}
S=&-A M_{P}^{2}\int d^{4}x 2a[H\Phi+\dot{\Psi}]
\partial_{i}N_{i}(1+2\eta\bar{F})-4 A \eta M_{P}^{2}\int d^{4}x a H
N_{i}\partial_{i}\delta F\\
&-2\eta\Upsilon M_{P}^{2}\int d^{4}x a[N_{i}\partial_{i}\dot{\delta
F}-\partial_{i}\Phi N_{i}\dot{\bar{F}}-3H N_{i}\partial_{i}\delta
F]+ S_{Matter} \ .
\end{split}
\end{equation}
Further on,  performing the variation with respect to $N_i$ and
using the Fourier components of the perturbations we obtain
\begin{equation}\label{im3}
A(H\Phi+\dot{\Psi})(1+2\eta\bar{F})=\eta\big[\Upsilon\dot{\delta
F}-\Upsilon \Phi\dot{\bar{F}}+(2A-3\Upsilon)H\delta F\big]
-\frac{1}{2M_{P}^{2}}\delta q \ .
\end{equation}
In case of the variation of the action with respect to $\Phi$ we set
$N_{i}=0$ in the action and then expand the action up to the second
order in $\Phi$ and $\Psi$. This procedure, after some integration
be parts and using Eq. \eqref{gen11}, leads to
\begin{equation}
\begin{split}
\delta_{\Phi}S_{res}=\int d^{4}x a^{3} M_{P}^{2}\delta\Phi
\Big[&\Big(-12A\Phi H^{2}+\frac{4}{a^{2}}\partial^{2}\Psi
-12AH\dot{\Psi}-\frac{2\gamma_{1}}{a^{2}}\partial^{2}\Phi\Big)\times\\
&\times\Big(\frac{1}{2}+\eta\bar{F}\Big)
+\eta\big((12A-18\Upsilon)H^{2}-6\Upsilon\dot{H}-
\frac{2\Upsilon}{a^{2}}\partial^{2}\big)\delta F\\
&-\eta\dot{\bar{F}}\big(12\Upsilon H\Phi+6\Upsilon\dot{\Psi}\big) +
6\eta\Upsilon H\dot{\delta F}-\delta_{\Phi}S_{Matter}\Big] \
\end{split}
\end{equation}
so that we easily find the Fourier form of the equation of motion
for $\Phi$
\begin{equation}\label{im4}
\begin{split}
&\eta\big
[(6A-9\Upsilon)H^{2}-3\Upsilon\dot{H}+\frac{\Upsilon}{a^{2}}k^{2}\big]\delta
F
-3\Upsilon\eta\dot{\bar{F}}(2H\Phi+\dot{\Psi})\\
&+ 3\eta\Upsilon H\dot{\delta F}-\frac{\delta\rho}{2M_{P}^{2}}
=(3A\Phi H^{2}+\frac{k^{2}}{a^{2}}\Psi
+3AH\dot{\Psi}-\gamma_{1}\frac{k^{2}}{2 a^{2}}\Phi)(1+2\eta\bar{F})
\ .
\end{split}
\end{equation}
For reasons that will become clear later we derive $\delta p$ in two
different ways. In the first case we use \eqref{im1} and
\eqref{im4}. Then in order to eliminate $(\rho+p)$ in these formula,
we use \eqref{bae}. Collecting these pieces together we obtain an
equation
\begin{equation}\label{p1}
\begin{split}
\frac{\delta
p}{2M_P^{2}}=&\eta(2A\dot{\Psi}+4AH\Phi+\Upsilon\dot{\Phi}-\gamma
_1\frac{k^{2}}{3Ha^{2}}\Phi) \dot{\bar F}-2\eta H\dot{\delta
F}+2\eta\Upsilon\Phi\ddot{\bar
F}-\eta\Upsilon\ddot{\delta F}\\
&+\eta\big[(3\Upsilon-2A)\dot{H}+(9\Upsilon-6A)H^{2}
+\frac{2k^{2}}{3a^{2}}(A-2\Upsilon)\big]\delta F\\
&+\big[\dot{\Phi}H+2A\dot{H}\Phi+A\ddot{\Psi}+3A\Phi
H^{2}+3AH\dot{\Psi}+\frac{k^2}{3a^{2}}(\Psi-A\Phi)\\
&\hspace{.15cm}+(1-A)\frac{k^{2}}{3Ha^{2}}\dot{\Psi}-
\gamma_{1}\frac{k^{2}}{6a^{2}}(\frac{\dot{\Phi}}{H\Phi}+1)\Phi\big](1+2\eta
\bar{F}) \ ,
\end{split}
\end{equation}
where we also used the fact that
$ \dot{R}_\Upsilon\delta F=\delta R_\Upsilon\dot{\bar{F}} $.\\
The second way how to derive the relation for $\delta p$ is to use
eqs. \eqref{im2},\eqref{im3}. We again use eq.\eqref{bae} in order
to eliminate $(\rho+p)$, so that we obtain
\begin{equation}\label{p2}
\begin{split}
\frac{\delta
p}{2M_P^{2}}=&\eta[2\dot{\Psi}+2(1+A)H\Phi+\Upsilon\dot{\Phi}]
\dot{\bar F}-2\eta A H\dot{\delta F}+2\eta\Upsilon\Phi\ddot{\bar
F}-\eta\Upsilon\ddot{\delta F}\\
&+\eta\big[(3\Upsilon-2A)\dot{H}+(9\Upsilon-6A)H^{2}\big]
\delta F\\
&+\big[\dot{\Phi}H+2\dot{H}\Phi+\ddot{\Psi}+3\Phi
H^{2}+3H\dot{\Psi}\big](1+2\eta \bar{F}) \ .
\end{split}
\end{equation}
The right-hand side of eqs. \eqref{p1} and \eqref{p2} are the same
if
\begin{equation}\label{im5}
\begin{split}
&\Big[(1-A)\dot{\Phi}H+(1-A)\frac{k^{2}}{3Ha^{2}}\dot{\Psi}+\frac{k^{2}}{3a^{2}}(\Psi-A\Phi)-\frac{\gamma_{1}k^{2}}{6a^{2}}(\frac{\dot{\Phi}}{H\Phi}+1)\Phi\Big]\Big[1+2\eta
\bar{F}\Big]\\
&=2\eta\frac{k^{2}}{3a^{2}}(2\Upsilon-A) \delta
F+\eta\Big[2(1-A)H\Phi+2(1-A)\dot{\Psi}
+\gamma_{1}\frac{k^{2}}{3Ha^{2}}\Phi\Big]\dot{\bar{F}} \ .
\end{split}
\end{equation}
As is clear, the equations are simplified by taking $\lambda=1$,
which results in $A=1$. As a check,  note that for
$\Upsilon=1,\lambda=1$ and $\gamma_{1}=0$ these equations match the
corresponding relations for the usual $f(R/M_{P}^{2})$ gravity.
\subsection{ Tensor Metric Perturbations}
In this section we focus on the  metric tensor perturbations
$\gamma_{ij}$. Recall that the perturbed line element has the form
\begin{equation}
ds^{2}=-dt^{2}+a^{2}[\delta_{ij}+\gamma_{ij}]dx^{i}dx^{j},
\end{equation}
where $\partial_{i}\gamma_{ij}=\gamma_i^i=0$. In order to derive
equations in this sector we use the fact that $\sqrt{-g}$ does not
contain the  metric tensor perturbations up to the second order.
Further, using the traceless condition on $\gamma_{ij}$ it is easy
to show that the metric perturbations do not appear in $K$. Thus, it
turns out that the terms in \eqref{repla} which are proportional to
$\Upsilon$ and $\lambda$ do not contribute to the  metric tensor
perturbations. As a result the analysis of this sector is very
similar to the analysis of tensor fluctuations in standard $f(R)$
gravity. Explicitly, the second order action for these modes has the
form
\begin{equation}\label{t1}
\delta S|_{tensor}=\frac{M_{P}^{2}}{8}\int
d^{4}x[1+2\eta\bar{F}]\big[a
\gamma_{ij}\partial^{2}\gamma_{ij}+a^{3}\dot{\gamma}_{ij}^{2} \big]
\ .
\end{equation}
Then in order to avoid the ghost instability we must impose the
following condition
\begin{equation}\label{s1}
1+2\eta\bar{F}>0.
\end{equation}
Performing variation of \eqref{t1} with respect to $\gamma_{ij}$ and
using the following Fourier representation
\begin{equation}\label{1-26}
\gamma_{ij}=\int\frac{d^{3}k}{(2\pi)^{3/2}}
\sum_{s=\pm}\epsilon_{ij}^{s}(k)\gamma_{k}^{s}(t)
e^{i\overrightarrow{k}.\overrightarrow{x}} \ ,
\end{equation}
where $\epsilon_{ii}=k^{i}\epsilon_{ij}=0$ and
$\epsilon_{ij}^s(k)\epsilon_{ij}^{{s'}}(k)=2\delta_{ss'}$, leads to
the second order differential equation for $\gamma^s_k$
\begin{equation}\label{1-27}
\ddot{\gamma_{k}^{s}}+\dot{\gamma_{k}^{s}}\frac{d}{dt}\ln[a^{3}(1+2\eta\bar{F})]+(\frac{k}{a})^{2}\gamma_{k}^{s}=0
\end{equation}
which has the same form as in the standard $f(R)$-gravity.
\subsection{ Vector Metric Perturbations}
In this section we perform an analysis of vector metric
perturbations. As in the case of the prefect fluid we consider the
following form of the perturbed stress-tensor in the vector sector
\begin{equation}
\delta T^{0}_{i}|_{vector}=\delta q_{i}^{V} \ .
\end{equation}
Then from $\nabla_{\mu}T^{\mu\nu}=0$ it follows that
\begin{equation}
\dot{\delta q_{i}^{V}}+3H\delta q_{i}^{V}=0 \ .
\end{equation}
As in case of the metric perturbations the favorite gauge in this
sector is the so-called vector gauge which is defined by
\begin{equation}
ds^{2}=-dt^{2}+2aS_{i}dx^{i}dt+a^{2}\delta_{ij}dx^{i}dx^{j} \ ,
\end{equation}
where $\partial_{i}S_{i}=0$. Again, from the above definition and
the condition on $S_{i}$ it turns out that the terms in \eqref{ac1}
which are proportional to $\Upsilon$ do not contribute in this
sector. Then the second order action takes the following form
\begin{equation}
\delta S|_{vector}=-\frac{M_{P}^{2}}{2} \int d^{4}x
aS_{i}\partial^{2}S_{i}(1+2\eta\bar{F})- 2\int d^{4}x
a^{2}S_{i}\delta q_{i}^{V} \ .
\end{equation}
Then it is easy to see that the equation for $S_{i}$ in the Fourier
space has the form
\begin{equation}
M_{P}^{2}k^{2}(1+2\eta\bar{F})\frac{S_{i}}{a}=2\delta q_{i}^{V}
\end{equation}
which coincides with the equation for the vector perturbation in the
standard $f(R)$-gravity.
\section{ Some Phenomenological Implications}
\label{fourth} In this section we explore some phenomenological
implications of our results. In this section we take $A=1$ that
corresponds to $\lambda=1$. Note that this is a natural presumption,
since it is expected that $\lambda$ should approach to $1$ in the
low energy regime \cite{Horava:2008ih,Horava:2009uw}. It is
important to emphasize that the Einstein-Hilbert gravity predicts
$\Phi=\Psi$. In the more general case when $\Phi\neq \Psi$, it is
useful to define an anisotropic stress, $\delta\Sigma$ as
\begin{equation}
\Psi-\Phi=8\pi G a^{2}\delta\Sigma \ .
\end{equation}
Clearly a non-vanishing anisotropic stress is the signature of
modification of general relativity \cite{Ade2:2015lrj,silk}.
Further, it is also convenient to define
\begin{equation}\label{defgam}
\gamma\equiv\frac{\Psi}{\Phi} \ ,
\end{equation}
so that the Einstein-Hilbert gravity predicts $\gamma=1$.\\
But, from the Planck CMB temperature data, the current value for
$\gamma$ reported as \cite{Ade2:2015lrj}
\begin{equation}
\gamma_{0}-1=0.70\pm 0.94
\end{equation}
which is shown by yellow and gray colors in Fig. 1. On the other
hand if we consider the weak lensing data, we have
\cite{Ade2:2015lrj}
\begin{equation}
\gamma_{0}-1=1.36_{-0.69}^{+1.0}
\end{equation}
which is shown by the gray and blue colors in Fig. 1. \\
Let us now return to our model and consider the late-time cosmology
when we can neglect the radiation and the cold dark matter.
Explicitly, we consider two cases:
\begin{itemize}
\item \emph{\textit{Case I}}:
In this case we presume that the cosmological constant $\Lambda$
is non-zero while $ \eta=0 $. \\
The energy density of the cosmological constant is $ \rho=\Lambda
M_{P}^{2} $ so that \eqref{gen11} implies
\begin{equation}
H^{2}=\frac{\Lambda}{3} \ .
\end{equation}
Note that this result is the same as in the case of
Einstein-Hilbert gravity with the cosmological constant.\\
Since for the cosmological constant we have $\delta q=\delta
p=\delta\rho=0$, from eqs. \eqref{im3} and \eqref{im5} we have
\begin{subequations}\label{des23}
\begin{align}
\dot{\Psi}+H\Phi =0 \ ,\\
\Psi-\Phi=\frac{\gamma_{1}}{2}(\frac{\dot{\Phi}}{H\Phi}+1)\Phi \ .
\end{align}
\end{subequations}
If we combine these two equations together, we obtain
\begin{equation}
\ddot{\Psi}+\left(\frac{2}{\gamma_{1}}+1\right)
H\dot{\Psi}+\frac{2}{\gamma_{1}}H^{2}\Psi=0
\end{equation}
that has solution
\begin{align}\label{psires}
\Psi&=C_{1}e^{-Ht}+C_{2}e^{-\frac{2}{\gamma_{1}} Ht}
\hspace{.4cm}( \text{for} \ \gamma_{1} \neq2) \ ,\\
\Psi&=C_{1}e^{-Ht}+C_{2}Hte^{- Ht}\hspace{.3cm}( \text{for} \
\gamma_{1} =2) \ ,
\end{align}
where $ C_{1} $ and $ C_{2} $ are arbitrary constants. It is clear
that in order to avoid instability we have to require that
$\gamma_{1}>0$.\\
Using now eqs. \eqref{defgam}, \eqref{des23} and \eqref{psires}, we
obtain
\begin{equation}\label{fo1}
\gamma-1=\frac{\gamma_{1}}{2}
\Big[1-\frac{C_{1}e^{-Ht}+C_{2}\frac{4}{\gamma_{1}^{2}}
e^{-\frac{2}{\gamma_{1}} Ht}}{C_{1}e^{-Ht}+C_{2}
\frac{2}{\gamma_{1}}e^{-\frac{2}{\gamma_{1}} Ht}}\Big] \
(\hspace{.2cm} \text{for} \ \gamma_{1} \neq2)
\end{equation}
and
\begin{equation}
\gamma-1=\frac{C_2}{C_1-C_2+C_2 Ht} \ (\hspace{.2cm} \text{for} \
\gamma_{1} =2) \ .
\end{equation}
For the present Universe we have $H_{0}t_{0}\approx1$ and it is easy
to see that we can choose the parameters in the above equations to
obtain a consistent result with the reported data for
$\gamma_{0}-1$. Note also that the above relations show that if we
take $t\rightarrow\infty$, we have
\begin{eqnarray}
& & \gamma\rightarrow 1 \ , \quad\text{if}\hspace{.2cm}
\gamma_{1}\leq2 \ , \nonumber \\
& &\gamma\rightarrow \frac{\gamma_{1}}{2} \ , \quad \text{if}
\hspace{.2cm}\gamma_{1}>2 \ . \nonumber \\
\end{eqnarray}
\item \emph{\textit{Case II}}:
In the second case, we neglect the cosmological constant but $
\eta\neq0$ so that we have only two independent equations which
determine $\Phi$ and $\Psi$. One of these equations is obtained by
multiplying eq.\eqref{im3} with $3H$ and then using the
corresponding result to eliminate some terms in eq.\eqref{im4}. This
procedure gives the following equation(for $A=1$)
\begin{equation}\label{des1}
\big[\frac{2k^{2}}{a^{2}}\Psi-\gamma_{1}\frac{k^{2}}{a^{2}}\Phi\big](1+2\eta
\bar{F})=-6\eta\Upsilon[\dot{\Psi}+H\Phi]\dot{\bar{F}}
+6\eta\Upsilon(\frac{k^{2}}{3a^{2}}-\dot{H})\delta F \ .
\end{equation}
Further, eq.\eqref{im5} takes the following form for $A=1$
\begin{equation}\label{des2}
\big[(\Psi-\Phi)-\frac{\gamma_{1}}{2}(\frac{\dot{\Phi}}{H\Phi}+1)\Phi\big]\big[1+2\eta
\bar{F}\big]\\
=2\eta(2\Upsilon-1)\delta
F+\frac{\eta\gamma_{1}}{H}\Phi\dot{\bar{F}}.
\end{equation}
In the de Sitter space, which is a very good approximation for the
late-time cosmology, eqs.\eqref{des1} and \eqref{des2} simplify
considerably
\begin{equation}\label{des1-1}
\big[2\Psi-\gamma_{1}\Phi\big](1+2\eta \bar{F})= 2\eta\Upsilon\delta
F
\end{equation}
and
\begin{equation}\label{des2-2}
\big[(\Psi-\Phi)-\frac{\gamma_{1}}{2}(\frac{\dot{\Phi}}{H\Phi}+1)\Phi\big]\big[1+2\eta
\bar{F}\big]\\
=2\eta(2\Upsilon-1)\delta F.
\end{equation}
If we combine these two equations, we obtain the following results
\begin{subequations}\label{des3}
\begin{align}
2(3\Upsilon-2)\Psi &=\Upsilon\gamma_{1}
(3-\frac{\dot{\Phi}}{H\Phi})\Phi-2(\Upsilon+\gamma_{1})\Phi\hspace{.4cm}
\left(\text{for} \ \Upsilon\neq0,\Upsilon\neq\frac{1}{2} \right) \ , \\
\Psi &=\frac{\gamma_{1}}{2}\Phi\hspace{.4cm}(\text{for} \Upsilon=0) \ ,\\
(\Psi-\Phi)&=\frac{\gamma_{1}}{2}\left(\frac{\dot{\Phi}}{H\Phi}+1\right)\Phi
\ , \hspace{.4cm}\left(\text{for} \Upsilon=\frac{1}{2}\right) \ .
\end{align}
\end{subequations}
Note that for $\Upsilon=2/3$, the first equation implies
\begin{equation}\label{dotPhiHPHi}
\frac{\dot{\Phi}}{H\Phi}=-\frac{2}{\gamma_{1}}.\hspace{.4cm}
\left(\text{for} \Upsilon=\frac{2}{3}\right) \ .
\end{equation}
\begin{figure}
\includegraphics{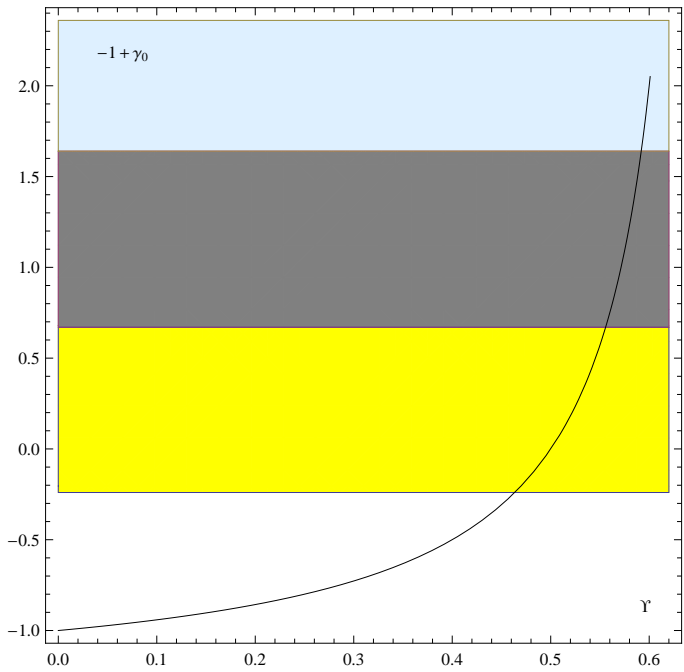}
\caption{\label{fig:epsart} Space of parameters ($ \gamma_{0}-1
$,$\Upsilon$) for the model. The yellow and gray regions show the
values for $\gamma_{0}-1$ from the Planck CMB temperature data
\cite{Ade2:2015lrj}. The gray and blue regions show the values for
$\gamma_{0}-1$ from the combination of the Planck CMB temperature
data and the weak lensing \cite{Ade2:2015lrj}. Solid line represents
the prediction of the model in the de Sitter space for
$\gamma_{1}\ll1$.}
\end{figure}
Finally when we insert \eqref{des3} and (\ref{dotPhiHPHi}) into eq.
\eqref{des1-1} or eq. \eqref{des2-2}, we obtain the equations for
the remaining variables that can be easily solved. Equations
\eqref{des3} are interesting results since they give a relation
between the parameters of the model and observation. If we take
$\gamma_{1}\ll1$ limit, it follows that
\begin{subequations}\label{des4}
\begin{align}
(3\Upsilon-2)\Psi &\approx-\Upsilon\Phi\hspace{.4cm}
\left(\text{for} \ \Upsilon\neq 0,\frac{1}{2},\frac{2}{3} \right) \ , \\
\frac{\Psi}{\Phi} &\ll 1\hspace{.4cm}(\text{for} \ \Upsilon=0) \ ,\\
\frac{\Psi}{\Phi}&\approx 1 \hspace{.4cm}\left(\text{for} \
\Upsilon=\frac{1}{2}\right) \ .
\end{align}
\end{subequations}
Note that, except for $\Upsilon=2/3$, in this limit( i.e
$\gamma_{1}\ll1$) the above relations do not depend on the form and
dynamics of $f$ in the action \eqref{ac1}. As  shown in Fig. 1, for
$\gamma_{1}\ll1$ one can choose $\Upsilon$ to reconcile the model
with the current observations.
\end{itemize}
\section{Conclusion}\label{fifth}
This short note is devoted to the analysis of the cosmological
fluctuations of the restricted $f(R)$-theory of gravity. We have
determined the background equations of motion and then we have
carefully analyzed the fluctuations above this background solution.
We show that the vector and tensor fluctuations have the same
dynamics as in case of the standard $f(R)$-gravity while the scalar
sector possesses new interesting possibilities which depend on the
values of coupling constants. In more details, we show that it is
possible to choose the values of these parameters so that the
predictions of  the restricted $f(R)$-gravity are in agreement with
recent observation data. In fact, we showed previously in
\cite{Chaichian:2015qjo} that the cosmological solutions found in
the restricted $f(R)$-gravity are in agreement with observation and
our current analysis of fluctuations confirms this fact as well.
This is by itself an  interesting result that suggests that it is
indeed plausible to consider theories with the restricted
diffeomorphism invariance as interesting alternatives to the fully
diffeomorphism invariant theories of gravity.

\vskip .5in \noindent {\bf Acknowledgement:}

 The work of J.K. was
supported by the Grant Agency of the Czech Republic under the grant
P201/12/G028.

\end{document}